\documentclass[11pt,twoside]{article} 
\usepackage{asp2004}
\usepackage{epsf}
\usepackage{psfig}
\usepackage{lscape} 

\begin{document} 
\title{Spectral Analysis of Super Soft X-ray Sources: \\ V4743 Sagittarii}
\author{T. Rauch,$^{1,2}$ M. Orio,$^{3,4}$ R. Gonzales-Riestra,$^5$ and M. Still$^6$} 
\affil{$^1$Dr.-Remeis-Sternwarte, Sternwartstra\ss e 7, 96049 Bamberg, Germany\\
       $^2$Institut f\"ur Astronomie und Astrophysik, Sand 1, 72076 T\"ubingen, Germany\\
       $^3$Istituto Nazionale di Astrofisica (INAF), Osservatorio Astronomico di Torino, 
           Strada Osservatorio, 20, 10025 Pino Torinese (TO), Italy\\
       $^4$Department of Astronomy, 475 N. Charter Street, University of Wisconsin, 
           Madison, WI 53706, USA\\
       $^5$XMM-Newton Science Operations Centre, Villafranca del Castillo, 
           P.O. Box - Apdo. 50727, 28080 Madrid, Spain\\
       $^6$NASA/GSFC, Code 662, Greenbelt, MD 20771, USA}
\begin{abstract} 
Half a year after its outburst, the nova V4743\,Sgr 
evolved into the brightest super-soft X-ray source in the sky
with a flux maximum around 30\,\AA, exhibiting resonance lines of
C\,{\sc v}, C\,{\sc vi}, N\,{\sc vi}, N\,{\sc vii}, and O\,{\sc vii}. 
We present preliminary results of an analysis of the  XMM-Newton RGS spectra
by means of NLTE model-atmosphere techniques.
\end{abstract}

\section{Introduction}
Classical novae occur in close binary systems (main-sequence star + white dwarf).
V4743\,Sgr has been discovered as a nova on Sep 20, 2002 (Haseda et al\@. 2002).
A first CHANDRA observation, taken with ACIS-S in Dec 2002 (Ness et al\@. 2003), 
shows a X-ray spectrum which is consistent with
a view on the optically thick atmosphere of an extremely hot white dwarf 
with strong C\,{\sc v}, N\,{\sc vi}, and N\,{\sc vii}
absorption features (indicating $T_\mathrm{eff}$ of 1 -- 2 MK, Ness et al\@. 2003)
which are blue shifted (\mbox{-2400}\,km/sec) --- 
most likely expanding gas which consists out of CNO-processed material.

\begin{figure}[ht]
\plotone{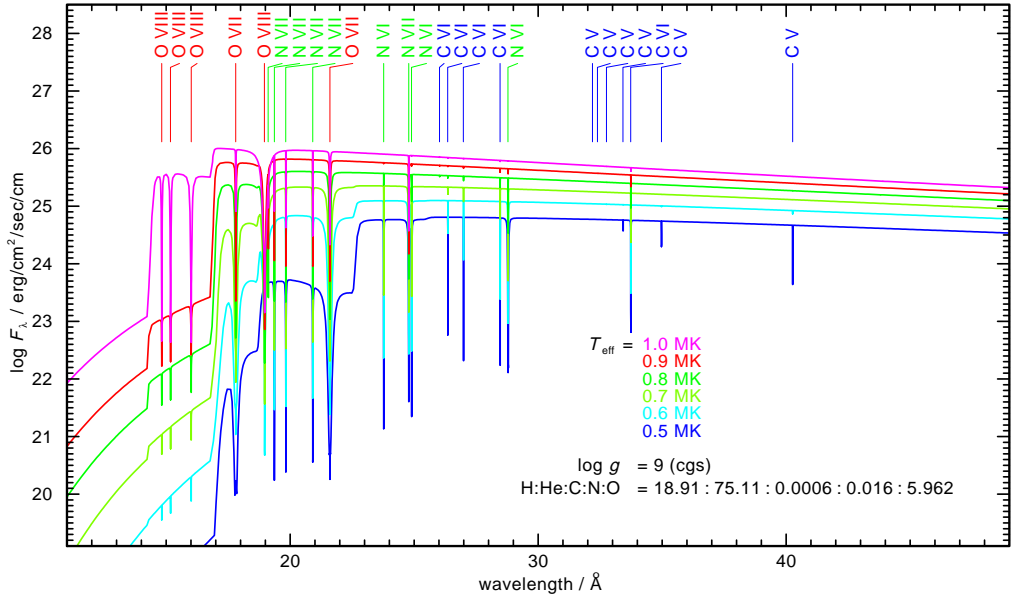}
\caption{Synthetic fluxes calculated from line-blanketed NLTE model atmospheres
including H, He, C, N, and O. The abundances are mass ratios.}
\label{synflux}
\end{figure}

\begin{figure}[hb]
\plotone{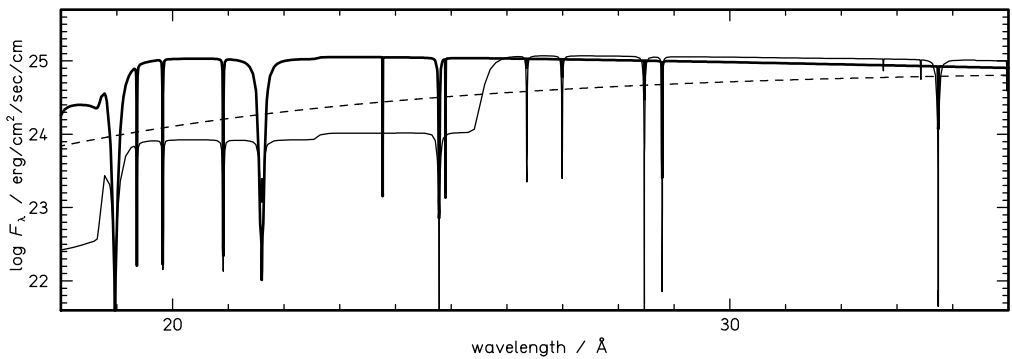}
\caption{Comparison of fluxes of $T_\mathrm{eff}$ = 700\,kK models 
with different abundance ratios (thick: like given in
Fig.\,\ref{synflux}, thin: solar, dashed: black body).}
\label{comflux}
\end{figure}

\section{Observations}
V4743\,Sgr was a very fast nova, with a steep decline in the optical light curve and large ejection
velocity. The time to decay by 3 mag in the visual ($t_3$), was 15 days 
and the Full Width at Half Maximum (FWHM) of the H$\alpha$ line reached 
2400 km/sec (Kato et al. 2002).

V4743\,Sgr was observed with the XMM-Newton satellite on Apr 4, 2003 for about 10\,h. The EPIC-pn count rate
in timing mode was 1348 cts/sec (Orio et al\@. 2003). The spectrum appears to be extremely
soft with absorption features (Fig\@. \ref{xmm}), which are attributed to the white dwarf's atmosphere.
Because of the high S/N achieved, we have used the XMM-Newton RGS spectra
obtained during this observation to start fitting the
stellar atmosphere, although we are also working at fitting
the previous spectrum, and later ones, obtained with CHANDRA LETG.

\begin{table}[t]
\caption{Statistics of the model atoms used in our calculations of the NLTE model atmospheres. 
  The notation is: NLTE = levels
  treated in NLTE, LTE = LTE levels, RBB = radiative bound-bound transitions}
\label{stat}
\smallskip
\begin{center}
{\small
\begin{tabular}{rlrrrrlrrr}
\tableline
\noalign{\smallskip}\noalign{\smallskip}
\multicolumn{1}{c}{atom} & ion & NLTE & LTE & RBB & \multicolumn{1}{c}{atom} & ion & NLTE & LTE & RBB \\
\noalign{\smallskip}
\tableline
\noalign{\smallskip}
H  & {\sc i}     &   5  &  11  &  10 & Ne & {\sc vii}   &  10  &  50  &  12 \\
   & {\sc ii}    &   1  &   -  &   - &    & {\sc viii}  &   8  &  18  &  15 \\
He & {\sc i}     &   1  &  21  &   0 &    & {\sc ix}    &   5  &   6  &   3 \\
   & {\sc ii}    &  10  &  22  &  45 &    & {\sc x}     &   1  &   0  &   0 \\
   & {\sc iii}   &   1  &   -  &   - & Mg & {\sc viii}  &   1  &  19  &   0 \\
C  & {\sc v}     &  29  &  21  &  60 &    & {\sc ix}    &   3  &  23  &   1 \\
   & {\sc vi}    &  15  &  21  &  26 &    & {\sc x}     &   2  &   3  &   1 \\
   & {\sc vii}   &   1  &   0  &   - &    & {\sc xi}    &   5  &   6  &   1 \\
N  & {\sc v}     &   5  &  15  &   6 &    & {\sc xii}   &   1  &   0  &   0 \\
   & {\sc vi}    &  17  &   7  &  33 & S  & {\sc x}     &   1  &   7  &   0 \\
   & {\sc vii}   &  15  &  21  &  30 &    & {\sc xi}    &   5  &  29  &   1 \\
   & {\sc viii}  &   1  &   0  &   - &    & {\sc xii}   &  10  &  12  &  15 \\
O  & {\sc vi}    &   1  &  34  &   0 &    & {\sc xiii}  &   9  &  21  &  10 \\
   & {\sc vii}   &  19  &   7  &  21 &    & {\sc xiv}   &   9  &   1  &  15 \\
   & {\sc viii}  &  15  &  30  &  30 &    & {\sc xv}    &   1  &   0  &   0 \\
   & {\sc ix}    &   1  &   0  &   - &    &             &      &      &     \\ 
\tableline
\noalign{\smallskip}\noalign{\smallskip}
   &             &      &      &     & \multicolumn{2}{r}{total} & 208 & 409 & 332 \\
\tableline
\end{tabular}
}
\end{center}
\end{table}

\section{NLTE Model Atmospheres}
The plane-parallel, static models are calculated with TMAP, the T\"ubingen NLTE Model Atmosphere Package 
(Werner et al\@. 2003, Rauch \& Deetjen 2003).
The first set of models is composed out of H+He+C+N+O, then Ne+Mg+S are added subsequently in
a line-formation calculation. However, in general TMAP can treat
all elements from hydrogen to the iron group simultaneously (Rauch 1997, 2003).
A small grid of model atmospheres 
was calculated in order to match the relevant temperature range 
(Fig\@. \ref{synflux}). The statistics of the model atoms which were used in the calculations are
summarized in Table \ref{stat}. 

\section{Results}
\begin{figure}[t]
\plotone{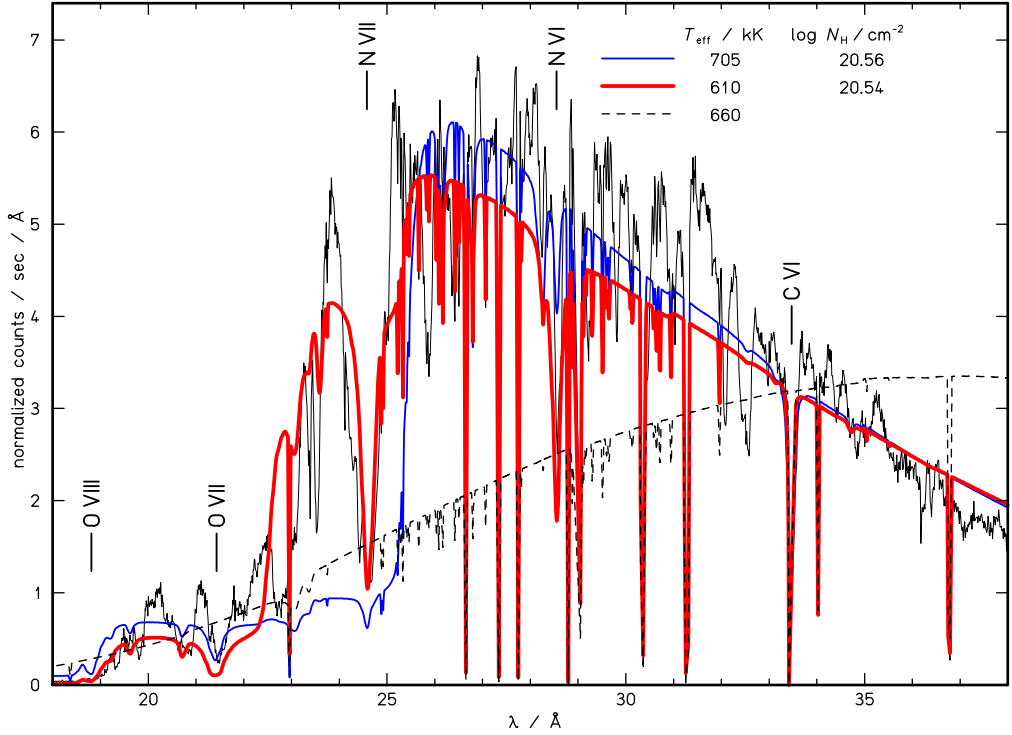}
\caption{Comparison of the XMM-Newton RGS-1 spectrum with
         NLTE model fluxes (thick: like given in
         Fig.\,\ref{synflux}, thin: solar, dashed: black body).
         $T_\mathrm{eff}$ and $N_\mathrm{H}$ are determined with XSPEC. 
         A black body (dashed) is shown for comparison only. 
         The synthetic spectra are shifted by $-$2400 km/sec.}
\label{xmm}
\end{figure}

The comparison of our preliminary model atmosphere fluxes with the Apr
2003 XMM-Newton RGS spectrum of V4743\,Sgr (Fig\@. \ref{xmm}) with the
help of XSPEC has shown that at solar abundance ratios, we cannot fit
the overall flux distribution because the flux at wavelength smaller
than 26\,\AA\ is too low (Figs.\,\ref{comflux}, \ref{xmm}). Although
the fit is not perfect yet, it is obvious that the C/N abundance ratio
is much smaller (about a hundred times) than the solar value, typical
for CNO processed material. The fit to the observation
(Fig\@. \ref{xmm}), based on our first test models
(Fig\@. \ref{synflux}), is best at a $T_\mathrm{eff}$ of about 610\,kK
and $N_\mathrm{H} = 3.5\cdot 10^{20}/\mathrm{cm^{-2}}$.

\clearpage

\acknowledgements{T.R\@. is supported by the DLR under grant 50\,OR\,0201.}


\begin{references}
\reference  Haseda, K., West, D., Yamaoka, H., \& Masi, 
G\@. 2002, IAU Circ\@. 7975
\reference  Kato, T., Hishikura, T., West, J.D., 
et al\@. 2002, IAU Circ\@. 7976
\reference  Ness, J.-U., Starrfield, S., Burwitz, V., 
et al\@. 2003, ApJ, 594, L127
\reference  Orio, M., Leibowitz, E., Rodriguez, P., 
et al\@. 2003, IAU Circ\@. 8131
\reference  Rauch, T\@. 1997, \aap, 320, 237
\reference  Rauch, T\@. 2003, \aap, 403, 709
\reference  Rauch, T., \& Deetjen, J.L\@. 2003,
            in: Workshop on Stellar Atmosphere Modeling,
            eds\@. I\@. Hubeny, D\@. Mihalas, K\@. Werner,
            The ASP Conference Series, 288, 103
\reference  Werner, K., Dreizler, S., Deetjen, J.L., Nagel, T., Rauch, T., \& Schuh, S.L\@. 2003,
            in: Workshop on Stellar Atmosphere Modeling,
            eds\@. I\@. Hubeny, D\@. Mihalas, K\@. Werner,
            The ASP Conference Series, 288, 31
\end{references}
\end{document}